\documentclass[sigconf,nonacm]{acmart}
\AtBeginDocument{%
  }
\setcopyright{none}
\renewcommand\footnotetextcopyrightpermission[1]{}
\copyrightyear{2026}
\acmYear{2026}
\acmConference[CSCW '26]{Broader Impacts of GenAI in Communication Workshop}{October 2026}{Salt Lake City, UT, USA}
\begin{document}

\title{Delegating Before Learning: Where Generative AI Sits in Students' Professional Communication}
\author{Jared Ren}
\email{renjared@uw.edu}
\affiliation{%
  \institution{Human Centered Design \& Engineering, University of Washington}
  \city{Seattle}
  \country{USA}
}

\author{Soobin Cho}
\email{soobin30@uw.edu}
\affiliation{%
  \institution{Human Centered Design \& Engineering, University of Washington}
  \city{Seattle}
  \country{USA}
}
\renewcommand{\shortauthors}{Ren and Cho}
\begin{abstract}
We conducted an interview study with twelve students on their use of generative AI in academic communication. Students delegated professional messages to AI most where the pressure to sound professional is highest: email to instructors and administrators. AI involvement ranged from correcting the writer's own text to working out and writing the message outright, and students checked AI-written text against two criteria: whether it looks like AI and whether it sounds like them. Building on these findings, we model the AI-mediated process of writing a student--instructor email at the highest level of involvement we observed, and compare it with an unaided model of writing the same messages, built from participants' accounts and a classic model of the writing process. Three differences emerge: the learning loop that builds writing skill is removed, the message is no longer written for its specific recipient, and the confidence a successful exchange returns goes to using the system rather than to the writer's own ability. From these differences we derive two risks, that individual capacities never form and that authenticity and trust in communication become work. Design can respond to both but is unlikely to be enough, so the risks also need research and policy attention.
\end{abstract}
\maketitle

\section{Introduction}
Generative AI is entering the messages students send. In AI-mediated communication, an intelligent agent modifies, augments, or generates messages on a communicator's behalf~\cite{hancock2020}.

Students' academic lives are full of these messages. They write to instructors, advisors, peers, and administrators across email, learning management systems, institutional platforms, and social media, and what these messages demand of the writer differs sharply by recipient.

In a prior interview study with twelve students, we found that AI use in academic communication concentrates on institutional and professional recipients and spans a wide range of involvement~\cite{jared}. That study closed with a question it could not answer: where does generative AI sit in the communication process?

This paper takes up the question by modeling the process itself, twice. We first model the AI-mediated process of writing an email to an instructor, at the highest level of AI involvement we observed. We then compare it with an unaided model of writing the same messages, built from participants' accounts and a classic model of the writing process.

We contribute (1) research findings on where and how students use generative AI for professional communication; (2) two process models of student--instructor email, with and without AI; and (3) three differences between the models, concerning learning, the recipient, and confidence. We conclude the paper by discussing two risks these differences imply, the design responses they invite, and why the risks also need research and policy attention.

\section{Where and How Students Use Generative AI}
We conducted two-phase interviews with twelve students recruited across multiple North American universities~\cite{jared}. Participants mapped their academic communication over the previous six months into recipient--channel groups, such as an instructor reached over school email, rated their AI use for each, and discussed selected groups in depth. This yielded 110 groups. The material here draws on a preliminary analysis of six of the twelve interviews.

\subsection{AI use concentrates on instructors}
Students used AI most with institutional and professional recipients. Of the 80 recipient--channel groups involving teaching staff or administrative and project-support recipients, 62 (77.5\%) reported AI use. For peers and peer groups, reached mostly through social media, 6 of 30 groups (20\%) did.

Participants' reasons cluster into three factors. \textit{Channel}: email is a professional channel and cannot be unsent, so it has to be right the first time. \textit{Recipient}: friends could be written to freely, while messages to instructors were written to make a professional impression. \textit{Message}: high-stakes content, such as requesting an extension or explaining a missed deadline, prompted AI use. Beneath all three sat doubt about their own writing ability: ``\textit{The emails are in general, just a professional way of conversing with someone. So in that sense is where I would use AI}'' (P11).

\subsection{AI involvement ranges from corrector to deliberator}
\begin{figure}[t]
  \centering
  \includegraphics[width=\columnwidth]{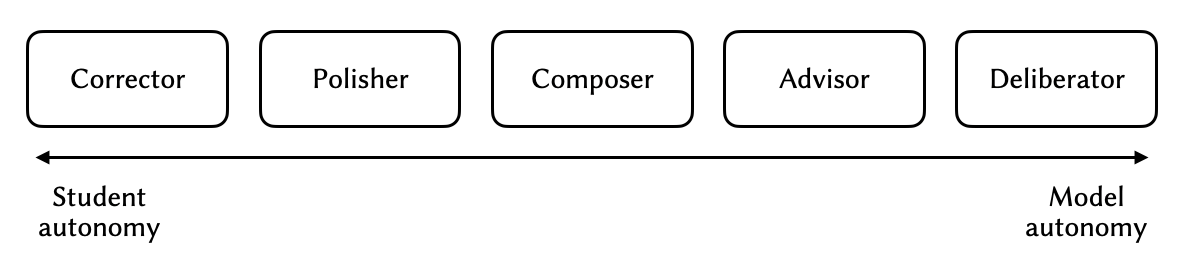}
  \caption{The roles participants described for AI in their messages, from corrector to deliberator. Autonomy shifts from the student to the model along the range; the AI-mediated model in this paper assumes the deliberator end.}
  \label{fig:range}
\end{figure}

AI involvement was not uniform (Figure~\ref{fig:range}). Participants described five roles for the system, from corrector at one end, through polisher, composer, and advisor, to deliberator at the other. The behaviors ranged from checking spelling and grammar, through adjusting tone and assembling a message from supplied points, to working out and writing the message outright. Autonomy shifts from the student to the model along the range. At some point within this range, the writer stops being the author whose text the system reviews and becomes the reviewer of text the system authored.

\subsection{Two checks on AI-written text}
When the system produced the text, participants evaluated it against two criteria: does it not look like AI, and does it sound like me. The first concerns the message's origin and passes when the reader cannot tell the text came from AI. The second is a voice judgment only the writer can make. Both checks structure the model that follows.

\section{Modeling AI-Mediated Student--Instructor Email}

\subsection{Scope}
We model the context where AI use was heaviest, email from students to instructors and teaching staff. The writer there is almost by definition a novice. Students are rarely fluent in professional academic register (the tone and form these messages are expected to take), the relationship is unequal, and the genre is rarely taught directly. Students without professional experience struggle to write such messages~\cite{huisprouse2023}. Email to administrative and project-support recipients behaved almost identically, so we treat this case as representative of professional academic email rather than a special one.

We assume the highest level of AI involvement we observed, the deliberator end of Figure~\ref{fig:range}, in which the system writes the message and the student revises. We choose it because the contrast is easiest to see there, and lower levels of involvement can be read as positions between the two models.

\begin{figure*}[t]
    \centering
    \includegraphics[width=0.9\textwidth]{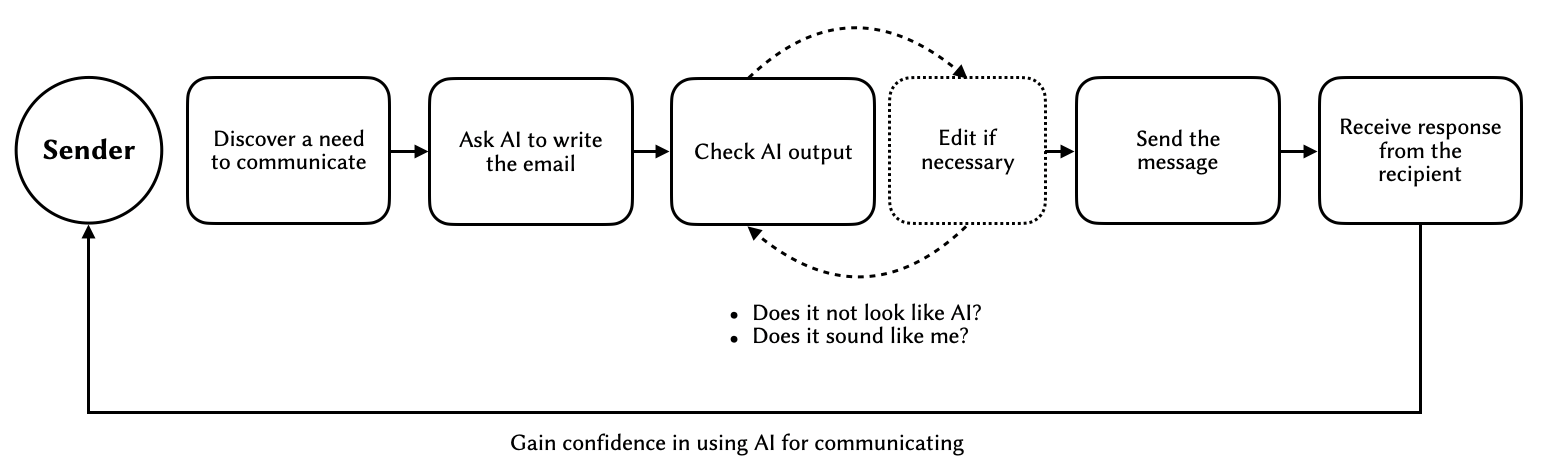}
    \caption{The AI-mediated process of writing an email to an
    instructor at the highest level of involvement we observed. The edit step is drawn with a dotted boundary because it is optional. The exchange returns to the writer as confidence in using the system rather than in their own ability to communicate.}
    \label{fig:ai_flow}
\end{figure*}

\subsection{The AI-mediated model}
In Figure~\ref{fig:ai_flow}, the writer discovers a need to communicate and asks an AI system to write the message. The prompt absorbs a step. Working out what to ask for and how to frame it, which otherwise happens before and during drafting, now happens inside the prompt, if at all. A prompt as short as ``ask my professor for an extension'' hands that work over too.

The writer then checks the output against the two criteria from Section~2.3 and edits if necessary. The edit step is drawn with a dotted boundary because it is optional and varied. Some participants sent the generated text unchanged. Others replaced only the sentences that did not sound like them or did not belong in a message to that recipient. Others swapped placeholders for real names and details. In no case was the message reworked as a whole. The message is sent, a reply arrives, and the outcome returns to the writer as confidence in using the system.

\section{An Unaided Model for Comparison}

\begin{figure*}[htbp]
    \centering
    \includegraphics[width=0.9\textwidth]{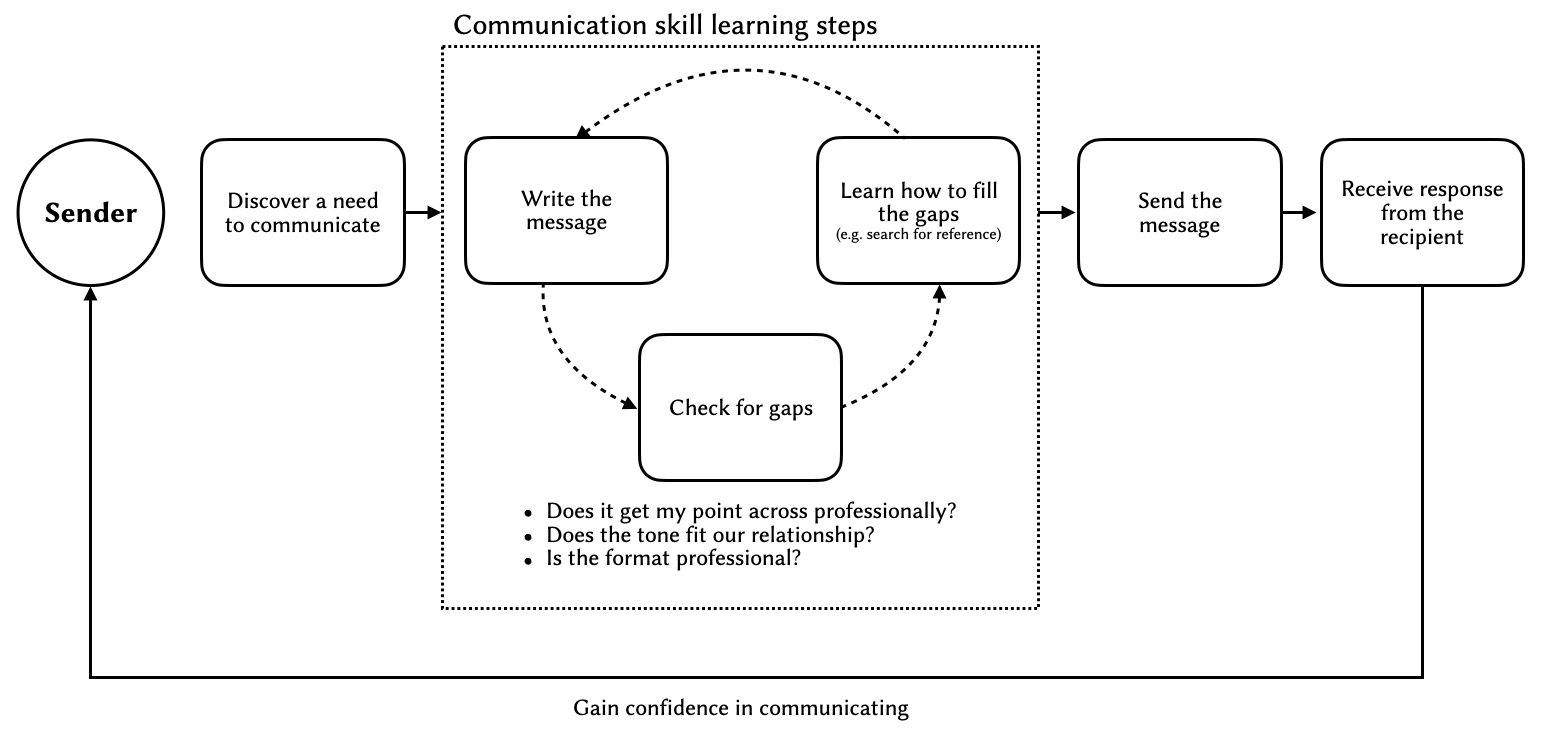}
    \caption{The unaided process of writing the same email. The inner loop is where the writer learns, since a failed check sends them to find out how to fill the gap and they return with genre knowledge they did not have. The exchange returns to the writer as confidence in their own ability to communicate.}
    \label{fig:unaided_flow}
\end{figure*}

Seeing what this process changes requires a model of the same message written without AI. Figure~\ref{fig:unaided_flow} presents one, built from how participants described composing such messages themselves and from the standard cognitive account of writing, in which planning, drafting, and reviewing loop back on one another rather than running in order~\cite{flowerhayes1981}.

The writer discovers the need, writes the message, and checks it for gaps: does it get the point across professionally, does the tone fit the relationship, is the format professional. When a check fails, the writer goes and finds out how to fix it, from an example, a template, a peer's message, or an email they once received and kept, and returns with something they did not have before. The loop turns until the message passes. The figure groups these steps as the communication skill learning steps. It is sent, a reply arrives, and the outcome returns to the writer as a small gain in confidence in their own communication.

\section{Three Differences}
Reading the models against each other shows three differences. The first two concern what the process no longer contains, and the third concerns what it returns.

\subsection{The learning loop is removed}
In the unaided model, a failed check sends the writer outward. Not knowing whether a format is right is what drives them to find an example, and each pass returns with something new. The loop yields a better message and a slightly better writer. In the AI-mediated model, a failed check produces a deletion. The writer cuts the sentence that does not fit, or fills in a real name, and moves on. Nothing turns the failure into an occasion to learn what would have fitted.

Steps also vanish at both ends of the process. The prompt absorbs planning, so working out the ask is never practiced. At the other end, some checks are dropped altogether. Writers learn, correctly, that the system does not misspell words or leave stray spaces, and stop looking for such errors. Only a narrow judgment survives, about whether a sentence sounds right for this writer and this reader. That is surface revision, and the surface is all that is left to revise, because the global decisions, what to say, in what order, in what register, were made elsewhere. Our participants could spot a professional email that works; they were not learning to write one.

\subsection{The recipient drops out}
The generated message is written for a generic professor. The actual recipient is this professor, with whom the student has a particular history, a degree of closeness, and a record of previous exchanges. Students' communication choices track exactly these features of the relationship~\cite{jared}, and pragmatics research on student--faculty email shows writers calibrating their requests to status and imposition~\cite{biesenbachlucas2007}. A system prompted for a professional email to a professor has none of them. It has a genre.

The swap is hard to notice because the output is competent writing addressed to no one in particular, a well-formed message for an average reader standing in for a real one. The edit pass catches the worst misfits, and participants did cut sentences that did not belong in a message to that recipient. But catching a misfit trims the message at its edges. Building the message around the relationship has to happen from the start.

\subsection{The confidence goes to the tool}
Both models close by returning the outcome of the exchange to the writer, but what returns differs. In the unaided model, a reply that goes well feeds the writer's sense of their own ability to communicate. In the AI-mediated model, it feeds their confidence in using the system.

Confidence in one's own ability grows most strongly through succeeding at the task oneself~\cite{bandura1997}, and success attributed to something outside oneself does not build it the same way. A student whose delegated email draws a warm, prompt reply has learned that the system writes good emails. The lesson stops there and never reaches their own ability. The skill that accumulates is real and useful, and it points at the wrong target. Writers become genuinely better at getting good messages out of the system, and no more able to produce one themselves. The opposite pattern appears in prior work, where students given support that scaffolds rather than generates report more confidence in their own writing afterward~\cite{huisprouse2023}.

\section{Risks and Responses}
In the short run, delegation almost certainly improves the messages students send. The risks lie elsewhere. We see two, and both can be read directly off the models.

The first risk is that individual capacities never form. Each step the AI-mediated model removes is a capacity the unaided model exercised. The prompt absorbs the decision-making and argumentation of working out what to ask for and how to frame it, the vanished search step removes information seeking, the missing draft removes writing practice, and the narrowed checks constrain independent, critical judgment of whether a message works. Capacities that are not exercised do not form, and the effect compounds. Low confidence in the genre leads to delegation, delegation takes away the practice that would build capacity and confidence, and the next message is delegated too. Our data supports the first step directly, as participants named low confidence as a reason for AI use. The second step is an inference from the model comparison and research on confidence, and testing it requires evidence that follows students over time, which we do not have.

The second risk is that authenticity and trust in the communication become work. The AI-mediated model adds two checks with no counterpart in the unaided model, and both concern how the message will be received rather than what it says. The voice check guards perceived sincerity, and the origin check makes sense only if being read as machine-written would lose the reader's trust. Passing both takes work. The writer must make the message sound like themselves and ensure it does not read as machine-written, labor that did not exist before and that is spent on where the message came from. The recipient side thins as well. The generated message is built for a generic professor, so the history and closeness that carry emotional connection survive only where the edit pass restores them. Once this labor is counted, whether delegation saves effort at all is open.

If the risks stem from the differences between the two models, design can try to put back what delegation removed. A system that treats a cut sentence as a signal, and shows the writer what would have fitted, restores the outward step of the unaided model. A system that asks who the message is for and what the relationship is, before drafting a word, builds the message around that relationship instead of the genre. Assistance that scaffolds rather than generates returns the confidence to the writer, and students using such systems report more confidence in their own writing afterward~\cite{huisprouse2023}.

Design alone is unlikely to be enough, because a design that teaches pays off across many messages while students choose their tools one message at a time. A tool that asks the writer to try first is more expensive at the moment of use than one that delivers the message ready to send, and every reason participants gave for delegating (the stakes, the permanence of email, the doubt in their own writing) weighs on the side of the ready message. A design that teaches has to survive that comparison each time a student faces a blank email.

This limit is why the risks need attention from research and policy as well. Neither risk will surface on its own. No single message is a problem. The student breaks no rule, the instructor is not wronged, and no vendor decision is to blame. Every visible sign improves: the emails are clearer, arrive faster, and are better received. The cost exists only in aggregate and over time, and it becomes visible only where someone deliberately looks. Research can look for it, since the models predict that heavy delegators should remain able to identify a professional email that works while becoming less able to produce one unaided. Policy can act on it, since institutions treat AI in student communication as an academic integrity question even though nothing in the process we model breaks a rule. Professional register, once acquired through ordinary daily practice, now has to be taught directly, and that is a curriculum question.

Where generative AI sits in the communication process is better answered by modeling the process twice than by locating AI within it once. The comparison shows that AI changes the process itself, and that its costs are invisible in the messages it produces. We offer it as a hypothesis.

% Discussion (and Conclusion) 

% Risk -> Design -> still don't think it would work -> need for additional research and policy

% There are n risks. 1) Impacts on individual capacities - critical thinking, decision-making, writing, argumentation, independent judgement, information seeking; 2) Authenticity, effort, and trust in communication, Perceived sincerity and emotional connection.

% If these risks are stemming from the differences between the two models, design can(?) put it back. [...]

% However, design is not sufficient because people won't use it. [...]

% So there's a need for research and policymaking beyond design. (This context is largely invisible and therefore unlikely to surface unless it is deliberately discussed. Yet the cost exists in aggregate and over time. We argue for its importance/urgency.)

\bibliographystyle{ACM-Reference-Format}
\bibliography{references}

@article{biesenbachlucas2007,
  author  = {Biesenbach-Lucas, Sigrun},
  title   = {Students Writing Emails to Faculty: An Examination of E-Politeness Among Native and Non-Native Speakers of English},
  journal = {Language Learning \& Technology},
  volume  = {11},
  number  = {2},
  pages   = {59--81},
  year    = {2007}
}

@article{flowerhayes1981,
  author  = {Flower, Linda and Hayes, John R.},
  title   = {A Cognitive Process Theory of Writing},
  journal = {College Composition and Communication},
  volume  = {32},
  number  = {4},
  pages   = {365--387},
  year    = {1981},
  doi     = {10.2307/356600}
}

@inproceedings{huisprouse2023,
  author    = {Hui, Julie and Sprouse, Michelle L.},
  title     = {Lettersmith: Scaffolding Written Professional Communication Among College Students},
  booktitle = {Proceedings of the 2023 CHI Conference on Human Factors in Computing Systems},
  series    = {CHI '23},
  year      = {2023},
  publisher = {ACM},
  address   = {New York, NY, USA},
  doi       = {10.1145/3544548.3581029}
}

@book{bandura1997,
  author    = {Bandura, Albert},
  title     = {Self-Efficacy: The Exercise of Control},
  publisher = {W. H. Freeman},
  address   = {New York, NY, USA},
  year      = {1997}
}

@article{hancock2020,
  title={AI-mediated communication: Definition, research agenda, and ethical considerations},
  author={Hancock, Jeffrey T and Naaman, Mor and Levy, Karen},
  journal={Journal of Computer-Mediated Communication},
  volume={25},
  number={1},
  pages={89--100},
  year={2020},
  publisher={Oxford University Press}
}

@inproceedings{jared,
author = {Ren, Jared and Cho, Soobin},
title = {"ChatGPT, Review This Message to My Prof" Investigating Students' Use of AI for Online Communication in Academic Settings},
year = {2026},
isbn = {9798400723780},
publisher = {Association for Computing Machinery},
address = {New York, NY, USA},
url = {https://doi.org/10.1145/3785651.3831512},
doi = {10.1145/3785651.3831512},
booktitle = {Proceedings of the 2026 Companion of the Computer-Supported Cooperative Work and Social Computing},
pages = {189–192},
numpages = {4},
location = {Salt Lake City, USA},
series = {CSCW Companion '26}
}
\end{document}